\documentclass[conference]{IEEEtran}
\IEEEoverridecommandlockouts
\usepackage{cite}
\usepackage{amsmath,amssymb,amsfonts}
\usepackage{algorithmic}
\usepackage{graphicx}
\usepackage{textcomp}
\usepackage{xcolor}
\def\BibTeX{{\rm B\kern-.05em{\sc i\kern-.025em b}\kern-.08em
		T\kern-.1667em\lower.7ex\hbox{E}\kern-.125emX}}
\begin{document}
	
	\title{Post-Quantum Cryptography\\
		%%{\footnotesize \textsuperscript{*}Note: Sub-titles are not captured in Xplore and should not be used}
		%%\thanks{Identify applicable funding agency here. If none, delete this.}
	}
	
	\author{\IEEEauthorblockN{ Pranjal}
			\IEEEauthorblockA{\textit{Pranveer Singh Institute of Technology}\\
			Kanpur, India \\
			pranjal10092005@gmail.com}

		\and
		
		\IEEEauthorblockN{ Atul Chaturvedi}
		\IEEEauthorblockA{\textit{Pranveer Singh Institute of Technology}\\
			Kanpur, India \\
			atul.chaturvedi@psit.ac.in}
		}
		%%\and
		%%\IEEEauthorblockN{2\textsuperscript{nd} Given Name Surname}
		%%\IEEEauthorblockA{\textit{dept. name of organization (of Aff.)} \\
		%\and
		%\and
		%\and
		%\and

	\maketitle
	
	\begin{abstract}
		In this survey we propose to cover the prose of post-quantum cryptography over classical cryptography. We talk about the various cryptographic methods that are being practiced to safeguard our information. The future of secure communication is expected to be the implementation of quantum-safe cryptographic systems, and that in the post-quantum era, the development of post-quantum cryptography is essential for ensuring the security of sensitive data.
	\end{abstract}
	%
	%\begin{IEEEkeywords}
	%component, formatting, style, styling, insert
	%\end{IEEEkeywords}
	\section{Introduction: }
		It is essential to understand the implications of advanced technologies such as quantum computing, as cyber security is becoming increasingly vital. Not just can quantum computing affect how businesses use quantum-resistant encryption to protect themselves, but it also has the potential to have a big impact on cyber security in the future. The advancement of quantum computing will profoundly affect cyber security, especially with regard to the state-of-the-art encryption techniques.
		Post-quantum cryptography (PQC) is the term for the new encryption techniques that researchers are creating and establishing to counter the threat posed by quantum computers. These new techniques are meant to be opposed to attacks from computers, both classical and quantum [1-12].

	\section{Cryptography: }
		The process of securing data and communications with codes so that only those who were meant to receive them is able to read and analyses them is known as \textbf{Cryptography}, preventing not wanted access to data in the process. The meaning of the prefix "crypt" is "hidden", and the suffix is "graphy" is "writing". The methods used in cryptography for protecting data are derived from mathematical concepts and a system of calculations based on rules, or algorithms, which transform communications into kinds of data that are difficult to understand. These algorithms are used to generate cryptographic keys, digitally sign documents, verify data privacy, browse the internet, and protect private transactions like debit and credit card purchases.\\
		In the modern era of computer science, cryptography is frequently linked to the process of converting a regular plain text to a \textbf{cypher text} "A text created so that only the person who intended it can decode it", this procedure is known as \textbf{encryption}. The process of translating cypher text into plain text is known as \textbf{Decryption}. 
	\subsection{Importance: }
		The field of cryptography, which is concerned with the security of communication and information, comprises several fundamental elements that work together to guarantee the confidentiality, integrity, and legitimacy of data. We explore the basic building blocks that serve as the foundation for cryptographic systems in this investigation [14].
	\begin{itemize}
		\item \textbf{Confidentiality:} It allows you to secure your data and information in such a way that only correct person should excess the document. It protects important data, such as personal messages, financial transactions and login credentials, from unauthorized access.
		\item \textbf{Integrity: } Hash functions in cryptography are used to confirm the accuracy of data. A piece of data can be uniquely hashed to identify any changes made to it and confirm that it wasn't tampered with throughout transmission or preservation.
		\item \textbf{Authentication: } Mechanisms for confirming the identities of those communicating are provided by cryptography. By doing this, impersonation is less likely to occur and messages are only shared between reliable parties.
		\item \textbf{Non-repudiation: } Digital signatures and other cryptographic techniques offer a way to authenticate an online communication or transaction. This keeps people from denying their involvement and aids in establishing accountability.
		\item \textbf{Secure transactions: } The security of online financial transactions and e-commerce depends strongly on cryptography. It guarantees the privacy and security of sensitive data, including credit card numbers, when making purchases online [14].
		\item \textbf{Data Protection: } Data kept on servers, databases, and other devices is encrypted using cryptography. This keeps the data safe even in the event of illegal access because it is encrypted and cannot be read without the proper decryption key.
		\item \textbf{National Security: } Governments and armed forces employ cryptography to protect private data pertaining to national security. It assists in protecting communication channels and classified data.
		\item \textbf{Privacy: } To protect people's privacy in messaging, email, and social media, among other online interactions, cryptography is crucial. Users can converse and exchange data without worrying about being secretly watched.
		\item \textbf{Trust: } A fundamental component of building trust in the digital sphere is cryptography. It promotes the expansion of digital services and communication by giving users peace of mind about the security and privacy of their online interactions.
		\item \textbf{Cyber-Security: } An essential part of cybersecurity is cryptography, which aids in defending enterprises against online dangers like identity theft, data breaches, and other types of cyber-attacks.
	\end{itemize}
	
	\section{Cryptographic Algorithms: }
		In today's world many different types of cryptographical algorithms are widely used in many social media platforms and banking services etc. Some of them are :
	\subsection{Symmetric-Key Algorithm: }
		In this encryption system, messages are encrypted and decrypted using a single common key shared by both the sender and the recipient [15]. Although symmetric key systems are quicker and easier to use, there is a drawback sender and recipient must exchange keys in a secure way. The \textbf{Data Encryption System }(DES) and the\textbf{ Advanced System} (AES) are the two most widely used symmetric key cryptography systems [15]. The types of Symmetric algorithms are as:\\
		Advanced Encryption Standard (AES)\\
		Data Encryption Standard (DES)\\
		Triple Data Encryption Standard (3DES)\\
	 	Blowfish 
%	\begin{figure}
%		\centerline{\includegraphics[scale=.7]{Symmetrickey.png}}
%		\caption{Symmetric Key}
%		\label{fig}
%	\end{figure}
	
	\subsection{Asymmetric-Key Algorithms (Public-Key Cryptography): }
		Data is encrypted and decrypted using two related keys in asymmetric key algorithms, commonly referred to as public-key cryptography. The keys have mathematical relationships, every key has a distinct purpose.
		One key is used to encrypt data, and a different key is used to decrypt the data. A public key is a cryptographic key that can be used by anyone to encrypt a message. Only the intended recipient can decrypt the message with their private key [16].\\
		This type of algorithm is used in :\\
		\textbf{RSA (Rivest-Shamir-Adleman):} Usually used for digital signatures, safe key exchange, and small-scale data encryption.\\
		\textbf{Elliptic Curve Cryptography (ECC):} Strong security with shorter key lengths makes it appropriate for environments with limited resources, such as mobile devices.\\
		\textbf{Diffie-Hellman (DH):} Applied when parties want to securely exchange keys over an untrusted network.\\
%	\begin{figure}
%		\centerline{\includegraphics[scale=.3]{Asymmetrickey.png}}
%		\caption{Asymmetric Key}
%		\label{fig}
%	\end{figure}
	
	\subsection{Hash Function: }
		A mathematical function that accepts an input of any size and outputs a fixed-size result is known as a cryptographic hash function (CHF). The result is referred to as a hash code, hash digest, or hash value. CHFs have numerous uses in information security and are used to confirm the authenticity of data. Message authentication codes (MACs), digital signatures, and other authentication methods are a few examples [17]. \\
		Two types of Hash Function is used: \\
		\textbf{SHA-256, SHA-3 (Secure Hash Algorithm):} Part of the SHA-2 and SHA-3 families, these hash functions are commonly used for data integrity verification.\\
		\textbf{MD5 (Message Digest Algorithm 5):} Although considered insecure for cryptographic purposes, MD5 is still used in some non-security-critical applications [17].
	\subsection{Digital Signatures: }
		The public-key building blocks of message authentication are digital signatures. Handwritten signatures are frequently used on handwritten or typewritten messages in the real world. They serve to bind the signatory to the communication.
		A digital signature, on the other hand, is a method that links an individual or organization to digital data. The recipient and any other third party may independently verify this binding [18].
		The digital signature is a cryptographic value that is determined using the signer's secret key and the data [18].
%		In the real world, the recipient of a message needs to be guaranteed that it is from the sender and that he cannot deny its source.\\
%	\begin{figure}
%		\centerline{\includegraphics[scale=.4]{DigitalSignature.jpg}}
%		\caption{Digital Signature}
%		\label{fig}
%	\end{figure}
		There are mainly two types of algorithms in Digital Signature.\\
		\textbf{DSA (Digital Signature Algorithm):} Frequently employed for key exchange and digital signatures.\\
		\textbf{ECDSA (Elliptic Curve Digital Signature Algorithm):} A digital signature algorithm based on elliptic curves that provides robust security with shorter key lengths.
	
	\subsection{Key Exchange Algorithms: }
		In cryptography, key exchange algorithms facilitate the application of cryptographic algorithms. They enable the exchange of cryptographic keys over an unsecured channel, such as the internet, between clients and servers. This makes it possible for them to conduct secure electronic transactions. It uses some mathematical formulas to generate the respective keys [19].
		There are mainly two algorithms for this method of cryptography:\\
		\textbf{Diffie-Hellman Key Exchange (DHKE):} It permits the safe agreement of two parties over an untrusted network on a shared secret.\\
		\textbf{ECDH (Elliptic Curve Diffie-Hellman):} An elliptic curve cryptography-based Diffie-Hellman variant.
		
	\subsection{Secure Communication Protocols: }
		Cryptography allows you to communicate secretly in the presence of malicious third parties. when two individuals talk to each other without someone else hearing them. The entities must communicate in a way that is impervious to eavesdropping or interception in order to accomplish this. 
		For secure communication we use below algorithms:\\
		\textbf{TLS (Transport Layer Security):} Used to protect online communication; frequently uses cryptographic algorithms such as RSA or ECC for digital signatures and key exchange, and AES for encryption.\\
		\textbf{SSH (Secure Shell):} Makes use of a number of cryptographic techniques to transfer files and enable secure remote access.
		
	\section{Threats and Solutions: }
		The area of cryptography is constantly changing, and new technologies bring new risks to established cryptographic methods. Here are some possible concerns and threats for current cryptographic techniques in the future, along with potential remedies [20].
	\subsection{Quantum Computing: }
		\textbf{Threats:} Popular public-key cryptography algorithms like RSA and ECC could be broken by quantum computers because they are capable of solving specific mathematical problems quickly.\\
		\textbf{Solution:} Research on post-quantum cryptography is ongoing, with the goal of creating algorithms that are safe even when quantum computers are present. As possible substitutes, algorithms like hash-based, code-based, and lattice-based cryptography are being investigated.
		
	\subsection{Side Channel Attacks: }
		\textbf{Threats:} Side-channel attacks use timing data or power consumption leaks that occur during computation to deduce secret keys.\\
		\textbf{solutions:} Side-channel attacks can be decreased by putting countermeasures like physical security measures, secure coding techniques, and constant-time algorithms into place. In addition, it is essential to design cryptographic algorithms that are resistant to side-channel attacks.
	
	\subsection{Cryptanalysis and Algorithmic Breakthroughs: }
		\textbf{Threats:} Developments in cryptanalysis could result from increases in computational power and mathematical research, making even the most secure algorithms vulnerable.\\
		\textbf{Solutions:} It is crucial to update cryptographic standards and algorithms on a regular basis in accordance with the most recent research findings. A multifaceted and cooperative cryptography community can aid in the prompt identification and remediation of weaknesses.
		
	\section{Post Quantum Cryptography: }
		The term "post-quantum cryptography" describes cryptographic methods and algorithms intended to maintain security throughout the quantum computing era. The security of commonly used cryptographic algorithms may be compromised by quantum computer's potential to solve some mathematical problems far more quickly than classical computers, especially if those algorithms are based on problems that are challenging for classical computers but simple for quantum ones [21].
	\subsection{How do Post Quantum Helps us: }
		To reduce the threats of cryptographic system post-quantum cryptography is essential. Adopting post-quantum cryptography proactively reduces the risks connected to the ultimately development of powerful quantum computers.
		The key aspects and goals are as:\\
		\textbf{Quantum-Safe: }Creating cryptographic algorithms that are resist the attacks by quantum computers. Many times, these algorithms are based on mathematical puzzles that are difficult to solve even for quantum computers [21].\\
		In this we uses the following algorithms:
		\begin{itemize}
			\item NTRU	(Lattice-based cryptography)
			\item CRYSTALS-Dilithium	(Shortest Integer Solution)
			\item Falcon	(Shortest Integer Solution)
			\item SPHINCS+	(hash based)
		\end{itemize}
		\textbf{Public-Key Cryptography Replacement:} Popular public-key cryptographic algorithms, like RSA and ECC (Elliptic Curve Cryptography), depend on mathematical issues that could be effectively resolved by quantum computers, such as integer factorization and elliptic curve discrete logarithm. These weak public-key cryptography algorithms are intended to be replaced by post-quantum cryptography algorithms.Most PQC algorithms will use a larger key size. For example, AES with keys greater than today's 128-bit keys.\\
		\textit{Elliptic Curve Cryptography (ECC)} based on the mathematics of elliptic curves, rather than the factorization of large prime numbers used in RSA.\\
		\textit{Diffie-Hellman (DH)} a key-agreement algorithm, rather than an encryption/decryption algorithm like RSA.
		
		\textbf{Quantum Key Distribution (QKD):} It is a class of cryptographic protocols that allows two parties to exchange keys securely by utilizing the ideas of quantum mechanics. QKD offers a secure key exchange mechanism that can withstand potential quantum attacks, it is regarded as quantum-safe [21].\\
		Some examples of QKD protocols are the following:
		\begin{itemize}
			\item BB84
			\item Silberhorn
			\item Decoy state
			\item KMB09
			\item E91
		\end{itemize}
		\textbf{NIST Post-Quantum Cryptography Standardization:} For post-quantum cryptography, the National Institute of Standards and Technology (NIST) started the standardization process in 2016. In the post-quantum era, NIST is actively testing and standardizing new cryptographic algorithms that may be applied to a range of security applications.\\
		NIST invited international cryptography specialists to submit potential algorithms for the project in 2016. As of the deadline in November 2017, 69 eligible algorithms had been submitted by experts from various countries. \\
		As the outcome of the (PQC) Standardisation Process, NIST chose four algorithms : CRYSTALS–KYBER, CRYSTALS–Dilithium, FALCON, and SPHINCS+ for standardisation in August 2023.\\
		For the majority of use cases, NIST will suggest CRYSTALS-KYBER (key establishment) and CRYSTALS-Dilithium (digital signatures) as the two main algorithms [15-16]. \\
		\textbf{Hybrid Cryptography:} In the meantime until quantum computers become widely accessible, hybrid Cryptography are frequently suggested. By fusing post-quantum and classical Cryptography algorithms, these systems create a link between modern security procedures and those needed in the post-quantum era. In the properties of
		symmetric and Asymmetric key system is fused to make a hybrid sysystem that will secure the large amount of data.\\
		Here's how hybrid cryptography works for secure file storage:
		\begin{itemize}
			\item A file is first encrypted using a symmetric key.
			\item Then file is further encrypted using the recipient's public key algorithms.
			\item By using a private key to decrypt the symmetric key, the recipient can then make sure that the file is only accessible to those who are authorized.  
		\end{itemize}
	
	\section{Conclusion: }
	The emergence of post-quantum cryptography represents a critical and early response to the approaching threat posed by quantum computers to traditional cryptographical systems. As quantum computing technology advances, the weaknesses of widely used encryption algorithms become increasingly evident. The research and development of post-quantum cryptographic solutions aim to create robust and secure methods that can withstand the computational power of quantum machines.
	
	PQC is an advantageous necessity for safeguarding our digital future, not just an advancement in technology. In an era where quantum computers could potentially compromise current encryption methods, governments, industries, and individuals must work together in order to implement these fresh cryptographic standards, making the protection of private data.
	
	The need to get ready for quantum-resistant solutions is evident, even though post-quantum cryptography is still developing and is the subject of continuous research and consistency efforts. Organizations investing in post-quantum cryptography algorithms must adopt a comprehensive strategy to guarantee a seamless transition without sacrificing existing security protocols.
	Our commitment to staying ahead of technological threats and protecting the confidentiality and integrity of our digital communications is shown by the path towards PQC. Considering the difficulties posed by quantum computing, it is a crucial step towards constructing a strong and secure foundation for information security in the future.

\end{document}